\documentclass[11pt,twoside]{article}

\usepackage{asp2006}
\usepackage{epsf}
\usepackage{psfig}
\usepackage{lscape}

\begin{document}

\title{The Origin of the soft excess in AGN}   

\author{Chris Done, Marek Gierlinski, Malgosia Sobolewska, Nick Schurch}   

\affil{Physics Department, University of Durham, Durham DH1 4ED, UK}    

\begin{abstract} 

We discuss various ideas for the origin of the soft X--ray excess seen
in AGN. There are clear advantages to models
where this arises from atomic processes in partially ionised rather
than where it is a true continuum component. However, current data
cannot distinguish between models where this material is seen in
reflection or absorption. While higher energy data may break the
degeneracies, we also suggest that strong outflows are 
extremely likely to be present, lending more physical plausibility to
an absorption origin. This more messy picture of NLS1's means that
they are probably not good places to test GR, but they do give insight into the
spectra expected from the first QSO's in the early Universe. 

\end{abstract}


\section{Introduction}

Many AGN show X--ray spectra which rise smoothly below 1~keV above the
extrapolated 2--10~keV emission. This soft X-ray excess is generally
fairly featureless. It cannot be resolved into a series of emission
lines with gratings, though there are some discrete
emission/absorption features superimposed on it. Thus it
looks like an apparent continuum component, but it is at far too high
a temperature to be simply the high energy tail of the accretion disc
emission. Fig 1a shows the soft excess in PG1211+104 
which is at much higher energies than predicted from the
expected accretion disc spectrum for the best estimates of mass and
$L/L_{Edd}$ for this object (Gierlinski \&
Done 2004).

The next most obvious continuum origin for the soft excess is as
Compton scattered disc emission. The observed rollover at $\sim
0.6$~keV implies an electron temperature of $kT_e\sim 0.2$~keV, and
the shape of the spectrum implies a large optical depth $\tau\sim 20$
(Gierlinski \& Done 2004). This low temperature, high optical depth
Comptonisation region is in addition to the high temperature,
low optical depth Comptonisation which makes the 2--10~keV component,
extending up to high energies. However, while this fits the shape
of the soft excess (Fig 1a), it predicts that it should relate to
the disc emission as this provides the source of seed photons for the
low temperature Comptonisation. A survey of PG Quasars in the
XMM--Newton database shows that this is not the case. All these
objects show a soft excess {\em with very similar temperature} of 
0.1--0.2~keV, yet span a range of over a factor 10 in disc
temperature. This makes it very unlikely that the soft excess is from
Comptonisation (Gierlinski \& Done 2004).

What then is the origin of the soft X--ray excess? The apparently fixed
temperature is much easier to explain if it arises from atomic rather than
continuum processes. One potential physical association is with the 
large increase in opacity between 0.7--3~keV due to OVII/OVIII and Fe
L shell absorption. This opacity jump from partially ionised material 
can produce a soft excess in two different geometries, one where the
material is optically thick and out of the line of sight, seen via
reflection (e.g. from an accretion disc). Alternatively, the material
can be optically thin and in the line of sight, seen in absorption
(e.g. a wind above the disc). We discuss each of these possiblities in
detail below.

\begin{figure*}
\begin{center}
\leavevmode
\psfig{figure=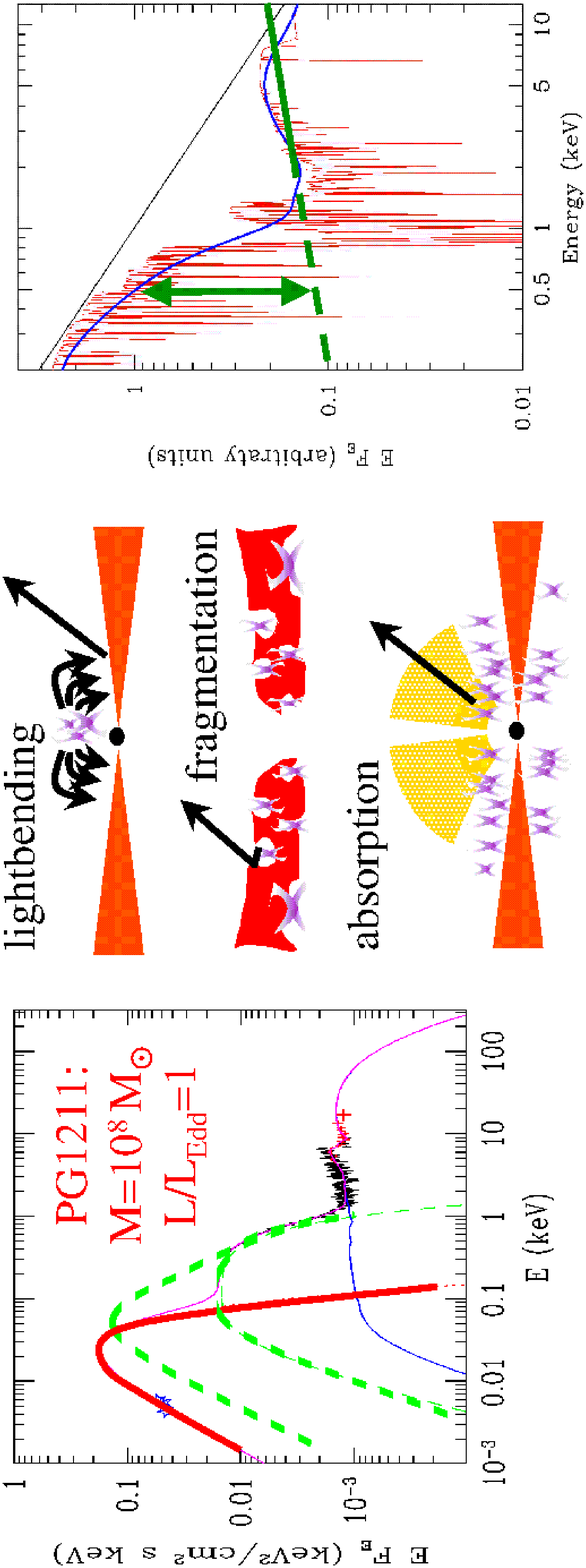,height=4.0cm,angle=-90} 
\vspace{-0.5cm}
\end{center}
\caption{a) The multiwavelength spectrum of PG1211. The EUV
shape of the soft excess is not constrained, but the soft X-ray
emission is much higher than predicted by the disc. b) shows 
possible reflection (lightbending and disc fragmantation) and
absorption geometries for the partially ionised material. c) shows the
characteristic absorption features are lost by doppler shifts so can
form the soft excess.}
\label{fig:sx}
\end{figure*}

\section{Reflection}

The reflection probablility is set by the balance between
photoelectric absorption and electron scattering. The higher the
photoelectric absorption, the less likely the photon is to be
reflected before being absorped so the lower the reflected
fraction. The increase in absorption opacity between 0.7--3~keV means
a decrease in reflection between these energies, and so to a big rise
in the reflected emission below 0.7~keV. This continuum reflection is
enhanced by emission lines from the partially ionised material,
especially OVII/VIII Ly $\alpha$ at 0.6--0.7~keV as well as Ly$\alpha$
lines from C,N and Fe L transitions (see e.g. Ross \& Fabian 2005).
However, while such partially ionised reflection spectra have long
been realised to match well to the energy of the soft X-ray excess
(e.g. Czerny \& Zycki 1994), the preponderance of sharp atomic
features in the models rules this out. If this is the origin of the
soft excess, then these features must be blurred into a
pseudo--continuum by strong velocity shear such as is produced by
reflection from a rotating disc (Cunningham 1975; Fabian et al
1989). Such models can match the shape of the soft excess, and
importantly the parameters required for the relativistic smearing to
make the soft excess can be the {\em same} as those required to
produce the associated iron K$\alpha$ line emission (Fabian et al
2002; 2004; 2005; Miniutti \& Fabian 2004; Crummy et al 2006, though
these also show that the highest signal--to--noise spectra require
{\em multiple} reflection components).

However, these parameters are uncomfortably extreme. The size
of the soft X-ray excess, parameterised as a ratio at 0.5~keV of the
data to the (extrapolated) power law fit to the 2--8~keV spectrum, can
be $>10$. Yet for isotropic illumination, the reflected emission below
1~keV cannot be larger than the illuminating flux, so limiting the
size of the soft excess to $<2$ (e.g. Sobolewska \& Done
2006).

The objects with these large soft X--ray excesses tend to be Narrow
Line Seyfert 1 galaxies. The large reflected fraction
could be produced in anisotropic illumination models, e.g. where the
X--ray source is extremely close to the black hole so that
lightbending suppresses the observed direct continuum flux and
enhances the disc illumination (Miniutti \& Fabian
2004). Alternatively, the disc might fragment into inhomogeneous
regions which hide a direct view of the intrinsic source flux (Fabian
et al 2002; 2005). These reflection geometries are sketched in Fig 1b.

As well as the reflected fraction being unexpectedly large, the amount of
relativistic smearing required is also extreme. Crummy et al (2006)
show that most of the PG Quasars require an inner disc radius which
implies an extreme spin black hole {\em and} an reflection 
emissivity which is much more centrally peaked than expected from
gravitational energy release. While this can be produced in the
lightbending scenario, this predicts that the objects with
largest reflection enhancement should have the strongest smearing
which is not always seen (Crummy et al 2006; Sobolewska \& Done 2006).

\section{Absorption}

The same physical process of the opacity increase between 0.7--3~keV
can also produce the soft excess
via absorption, plausibly from a wind from the accretion disc
(Gierlinski \& Done 2004, see Fig 1c). Again, relativistic velocity shear
is required to smear out the characteristic atomic features into a
pseudo--continuum but the difference between this and the reflection
model is that these motions are no longer Keplarian, so cannot be used
to simply infer the inner disc radius (and hence black hole spin). 

The absorption model fits the data from PG1211+104, a NLS1 with a
large soft X--ray excess, as well as the
reflection model. There is still some reflection present in the data,
but neither solid angle nor smearing are extreme.
However, for 1H0707-495, a NLS1
with an extremely large soft X--ray excess, there is also a strong 
sharp drop at 7~keV which cannot be well matched by the simple smeared
absorption (Sobolewska \& Done 2006). However, this could be an
artifact of the very simple velocity field (Gaussian!) assumed in the
model. A wind with coherent velocity structure produces P Cygni 
line profiles in resonance absorption lines such as He-- and H--like
iron K$\alpha$ at 6.7 and 6.95~keV. Including such profiles for the
iron line emission gives an excellent fit to the sharp drop at $\sim
7$~keV in 1H0707-495 (Done et al 2006). 

\section{Variability Behaviour}

Since both the reflection models and the absorption models can 
give comparable fits to the XMM--Newton spectra, we look instead at
variability properties in order to distinguish between them. 
The observed XMM--Newton {\it r.m.s.} variability spectra typically peak in the
0.7--3~keV region, strongly supporting the association of the soft
excess with atomic processes. In the reflection model this pattern of
variability can be produced by decomposing the spectrum into a
variable power law together with a more or less constant reflected
component. The higher fraction of the power law flux in the 0.7--3~keV
bandpass fits the observed peak in the {\it r.m.s.} variability (Ponti
et al 2006). It is hard to understand why
the reflected flux remains constant while the illuminating power law
varies, though again the lightbending model can reproduce this
behaviour (Miniutti \& Fabian 2004). By contrast, in the absorption
model, the variable power law flux induces correlated 
changes in the ionisation state of the absorber, $\xi=L/nr^2$. 
Higher
power law flux gives higher ionisation and less absorption in the
0.7--3~keV band, while lower illumination gives lower ionisation and
more absorption. These ionisation changes amplify the 0.7--3~keV 
variability, fitting the peaked {\it r.m.s.} variability spectra
(Gierlinski \& Done 2006). 

\section{Conclusions}

Neither spectra nor spectral variability in the 0.3--10~keV bandpass of
XMM--Newton can distingush between the reflection and absorption
origin of the soft X--ray excess. Suzaku data extending the bandpass 
beyond 10~keV may give the first direct test between the models as 
reflection predicts somewhat higher fluxes in the 10--50~keV bandpass
than absorption (Sobolewska \& Done 2006). 

More generally, we might have to use physical plausibility arguments,
especially as both reflection and absorption models are subject to
systematic uncertainties e.g. an unknown ionisation distribution with
distance for reflection and with velocity and column density and
covering factor for the associated emission in absorption (Sobolewska
\& Done 2006, Schurch \& Done 2006).  We {\em expect} that the NLS1
galaxies are messy, embedded in a powerful wind, as these are
typically high $L/L_{Edd}$ sources.  Physically it is unlikely that
these objects give us a clean view of the accretion disc, so their
spectra probably cannot be used for studies of General Relativity
without understanding more of the messy environment around a high
accretion rate black hole. Nonetheless, despite these difficulties, it
is very important to try to understand these spectra as they are the
local (high signal to noise) examples of the high mass accretion rates
expected in the early Universe from the first QSO's.



\end{document}